\documentclass[useAMS,usenatbib,usegraphicx]{mn2e}

\title[High-temperature behaviour of the $G$ ratio]{High-temperature behaviour of the helium-like K$\balpha$ $\bmath G$ ratio: the effect of improved recombination rate coefficients for calcium, iron, and nickel}
\author[J. Oelgoetz et al.]
  {J.~Oelgoetz,$^1$\thanks{E-mail: oelgoetz@lanl.gov}
  C.~J.~Fontes,$^1$ H.~L.~Zhang,$^1$ \newauthor M.~Montenegro,$^2$ S.~N.~Nahar,$^2$ and A.~K.~Pradhan$^2$ \\
  $^1$Applied Physics Division, Los Alamos National Laboratory, PO Box 1663, MS F663, Los Alamos, NM 87545, USA \\
  $^2$Astronomy Department, The Ohio State University, 140 W. 18th Avenue, Columbus, OH 43210, USA} 
\date{Released 2007 Xxxxx XX}

\pagerange{\pageref{firstpage}--\pageref{lastpage}} \pubyear{2007}
\usepackage[dvips,dvipsnames]{color}
\newcommand{\jqsrt}{J. Quant. Spectrosc. Radiat. Transf.}
\newcommand{\apj}{ApJ}
\newcommand{\aaps}{A\&AS}
\newcommand{\apjs}{ApJS}
\newcommand{\mnras}{MNRAS}
\newcommand{\jpbo}{J. Phys. B, At., Mol. Phys.}
\newcommand{\adndt}{At. Dat. Nuc. Dat. Tab.}
\newcommand{\aap}{A\&A}
\newcommand{\nat}{Nat}
\newcommand{\pre}{Phys. Rev. E}
\newcommand{\pra}{Phys. Rev. A}
\begin{document}

\setcounter{page}{0}
\thispagestyle{empty}
\begin{center}
\resizebox{\textwidth}{!}{\includegraphics*{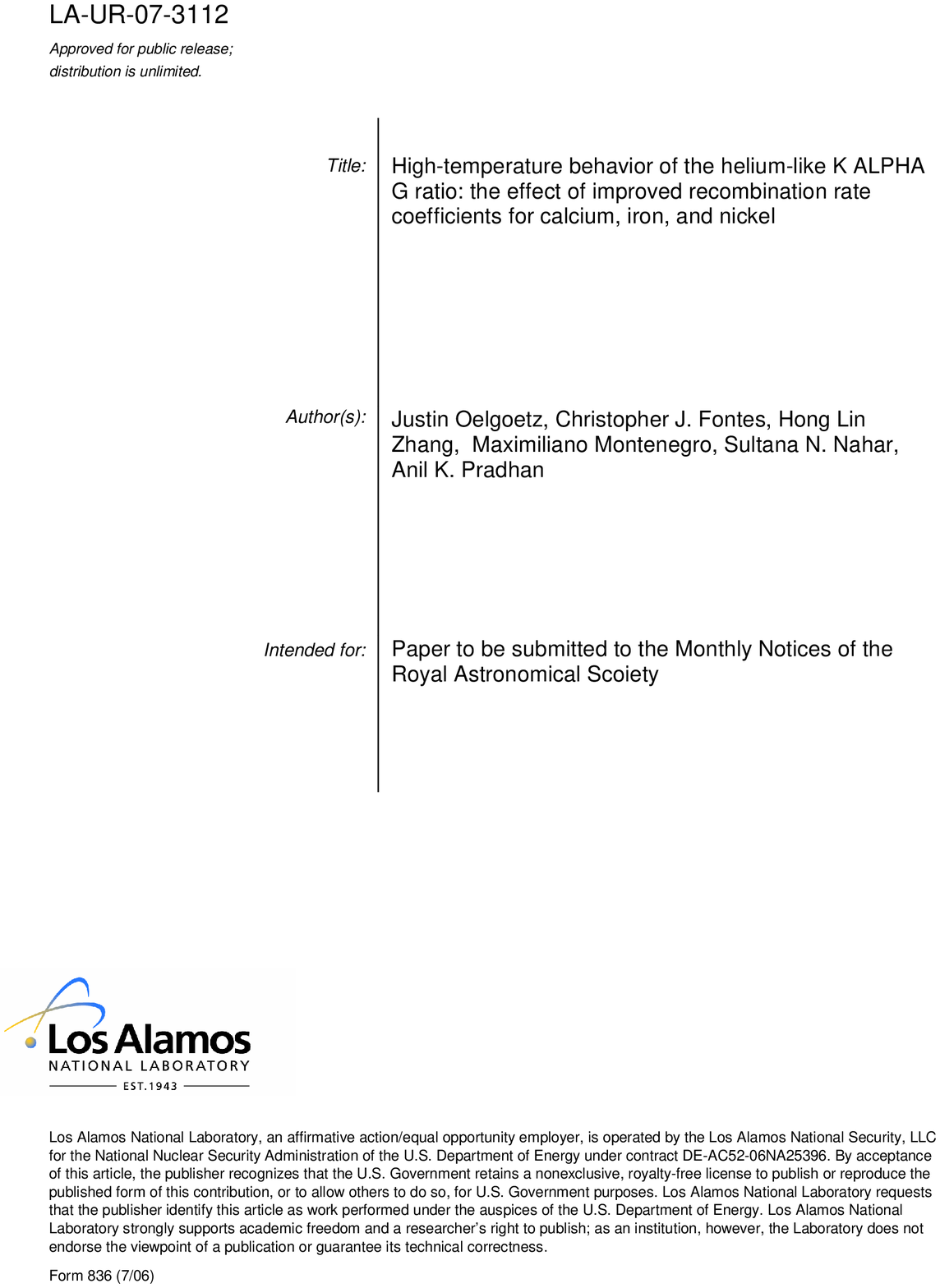}}
\end{center}

\maketitle
\label{firstpage}
\begin{abstract}
It is shown that above the temperature of maximum abundance, recombination rates into the excited states of He-like ions that are calculated using earlier, more approximate methods differ markedly from rates obtained from recent distorted-wave and R-Matrix calculations (unified recombination rate coefficients) for Ca, Fe, and Ni.  The present rates lead to $G$ ratios that are greatly lower than those resulting from the more approximate rates in previous works, by up to a factor of six at high electron temperatures.  Excellent agreement between the distorted-wave and the R-Matrix rates, as well as excellent agreement in the $G$ ratios calculated from them, provides support for the accuracy of these new values which have a broad applicability to the modelling and interpreting of X-ray spectra from a variety of astrophysical and laboratory sources.
\end{abstract}

\section{Introduction}
The K$\alpha$ emission lines of helium-like ions have been actively studied for quite some time, starting with \citet{Edlen-Tyren:1939}.  Much of the basic theory and interpretation that is now used for astrophysical spectroscopy was established by \citet{Gabriel-Jordan:1969-helines} in their work to interpret solar spectra.  While detailed discussions of the structure of the K$\alpha$ complex and line ratios can be found elsewhere \citep[see][]{Gabriel-Jordan:1969-helines,Gabriel-Jordan:1969-nature,Gabriel:1972-sats,Blumenthal-etal:1972,Mewe-Schrijver:1978stat,Mewe-Schrijver:1978nonstat,Pradhan-Shull:1981-helines,Pradhan:1982-helines}, in brief there are four He-like lines in the K$\alpha$ complex: $\mathrm{1s2p\;^1P_1\rightarrow1s^2\;^1S_0}$ (w), $\mathrm{1s2p\;^3P_2\rightarrow1s^2\;^1S_0}$ (x), $\mathrm{1s2p\;^3P_1\rightarrow1s^2\;^1S_0}$ (y), $\mathrm{1s2s\;^3S_1\rightarrow1s^2\;^1S_0}$ (z).  The `w' line is sometimes referred to as the resonance or dipole allowed line and `y' as the intercombination line as it is dipole allowed through fine-structure mixing.  The lines `x' and `z' both arise from higher order transitions, and as such are described by much slower rates.  The `x' line represents a magnetic quadrupole transition, and the `z' line represents a magnetic dipole transition.  \citet{Gabriel-Jordan:1969-helines} defined the temperature sensitive line ratio $G$ as
\begin{equation}
G=\frac{I(\mathrm{x})+I(\mathrm{y})+I(\mathrm{z})}{I(\mathrm{w})}\;,
\end{equation}
where $I(\mathrm{w})$ is the total emission in the `w' line in units of number of photons cm$^{-3}$ s$^{-1}$.  In general, the behaviour of the $G$ ratio is determined by the rate at which the He-like excited states are populated. At high temperatures these populations are determined by an interplay between the processes of recombination and impact excitation.  Recombination preferentially feeds the three lines arising from triplet states (`x', `y', and `z') while electron-impact excitation preferentially feeds the `w' line.  If the recombination rate is dominant, due, for example, to an increase in the number of H-like ions relative to the number of He-like ions, one would expect to see more flux in the `x', `y' and `z' lines than in the `w' line (as is observed in the case of recombining plasmas \citep{Pradhan:1985-cascade,Oelgoetz-Pradhan:2004}), resulting in a $G$ ratio greater than one.

For heavier atoms the situation is more complex as satellite lines produced by radiative decay from autoionizing three-electron states are a prominent part of the K$\alpha$ spectrum \citep[see][among others]{Gabriel-Jordan:1969-nature,Gabriel:1972-sats,Mewe-Schrijver:1978stat,Swartz-Sulkanen:1993,Bautista-Kallman:2000,Oelgoetz-Pradhan:2001}.  These satellite lines become less important as temperature increases.  As the high-temperature regime (above $10^7$~K) is the interest of this present work, no investigation of line ratios involving satellite lines has been made, although we note that satellite lines can be, and often are, important at temperatures well above $10^7$~K \citep{Oelgoetz-Pradhan:2001}.  If the spectra are not adequately resolved, these satellite lines should be included in any line ratios used for analysis.

As this study investigates the effect of recombination rates (both radiative (RR) and dielectronic (DR)) on the high-temperature behaviour of the $G$ ratio, a brief review of the data commonly used, how they are used, and the reported high-temperature dependence of the $G$ ratio is provided.  CHIANTI \citep{CHIANTIv1,CHIANTIv5}, XSTAR \citep{Bautista-Kallman2001-xstar}, and ATOMDB \citep{smith-etal-2001} are three commonly cited databases in astrophysics.  All three use different types of recombination data.

CHIANTI uses the recombination data of \citet{Mewe-etal.1985} for most helium-like ions which is based on the work of \citet{Burgess-Seaton-1960} on recombination into neutral helium.  The work of \citet{Mewe-Schrijver:1978stat} is also based on \citet{Burgess-Seaton-1960} and has been used in some recent investigations such as \citet{Oelgoetz-Pradhan:2001} as well as the earlier work by \citet{Pradhan-Shull:1981-helines}.   Fe and Ca are the exceptions to the above rule; for these elements CHIANTI uses the work by \citet{Bely-Dubau-etal.1982a,Bely-Dubau-etal.1982b} which is based on earlier studies of hydrogen by \citet{Burgess:1958}.  For completeness we note that \citet{Mewe-etal.1985} also depart from the above methodology and instead directly fit recombination data from \citet{Bely-Dubau-etal.1982a,Bely-Dubau-etal.1982b} for Fe XXV and Ca XIX.  The hydrogenic methodology of \citet{Bely-Dubau-etal.1982a,Bely-Dubau-etal.1982b} has also been used in calculations for other elements (e.g. \citet{Porquet-Dubau.2000}).

XSTAR takes a different approach; it calculates recombination rates from the photoionization cross sections produced by the Opacity Project \citep{Seaton-1987,TOPbase-1993}, using more approximate methods only for $\mathrm{n}>10$.  These Opacity Project cross sections were calculated using {R}-{M}atrix methods in {LS} coupling \citep{Burke-etal-1971,Berrington-etal.1987}.  Most recently \citet{Oelgoetz-Pradhan:2004} used unified recombination rates based on R-Matrix calculations \citep{Nahar-Pradhan:1994-urr} with relativistic fine structure included in the Breit-Pauli approximation.

ATOMDB uses the fits to distorted-wave photoionization data of \citet{Clark-etal-1986} for He-like cross sections.  These cross sections are integrated to obtain the corresponding rates in the standard manner.  Additionally there are numerous individual calculations that use recombination rates from distorted-wave data in a similar fashion.  Thus, much of the recombination data falls into three basic categories: data derived from more approximate methods (such as those by \citet{Burgess:1958} and \citet{Burgess-Seaton-1960}), data based on {R}-Matrix calculations \citep{Burke-etal-1971,Berrington-etal.1987}, and data based on distorted-wave calculations.

Both \citet{Mewe-etal.1985} and \citet{Bautista-Kallman:2000} calculate the emission intensities of Fe XXV as does \citet{Pradhan-Shull:1981-helines}, using the \citet{Mewe-Schrijver:1978stat} rate coefficients.  If one calculates $G$ ratios from the data presented in table IV of \citet{Mewe-etal.1985} the value of the $G$ ratio is steadily decreasing, with values of approximately 0.5, in between $10^8$ and $10^9$ Kelvin.  \citet{Pradhan-Shull:1981-helines} show a $G$ ratio taking a sharp upturn in the same temperature range and having a value greater than 1.0.  \citet{Bautista-Kallman:2000} for this same temperature range report a $G$ value in the range of 0.25 for Fe XXV, with a possibly slight rise as the electron temperature approaches $10^9$~K. Thus the disagreement among these calculations is significant and resolving this discrepancy is the goal of the present work.  To this end we use GSM \citep{Oelgoetz-Dissertation}, a code recently developed to model plasma emission spectra, to explore the effects of different types of recombination data on the $G$ ratio at high temperatures.   Data from recent distorted-wave and R-Matrix calculations are used and compared to rates calculated using expressions from \citet{Mewe-Schrijver:1978stat} and \citet{Mewe-etal.1985}; the exact rates in the XSTAR and ATOMDB databases are not considered here.  Additionally, only the effects of using different recombination data sets are explored while all other rates and the ionisation balance are identical for all models of a particular element.

\section{Theory}

GSM is a code based on the quasi-static approximation, that is the excited states of a given ion are considered to always be in equilibrium with the ground state of the ionisation stage from which they arise, and the neighbouring ionisation stages \citep[see][]{Loch-etal.2004}.  GSM makes the additional approximation that some states can be treated purely as conduits that, through radiative decay or auto-ionisation, allow population to transition into the final states of interest, thus neglecting the effect of collisions with these intermediate states.  Both approximations are valid in the case of low density plasmas where the radiation field can be neglected and are common to many of the previous calculations referenced in this work \citep[see][]{Gabriel-Jordan:1969-helines,Gabriel-Jordan:1969-nature,Gabriel:1972-sats,Blumenthal-etal:1972,Mewe-Schrijver:1978stat,Mewe-Schrijver:1978nonstat,Pradhan-Shull:1981-helines,Oelgoetz-Pradhan:2001,Oelgoetz-Pradhan:2004}.

As a consequence of these approximations the first step in any calculation is to solve for the total population of each ionisation stage.   This is done by solving the following set of coupled differential equations:
\begin{eqnarray}
\frac{dX_i}{dt}&=&N_e(X_{i+1}\alpha_{i+1 \rightarrow i}(\tilde{\varepsilon})+X_{i-1}C_{i-1 \rightarrow i}(\tilde{\varepsilon})) \nonumber \\ && + N_e^2(X_{i+1}\beta_{i+1 \rightarrow i}(\tilde{\varepsilon}) - X_{i}\beta_{i \rightarrow i-1}(\tilde{\varepsilon})) \nonumber \\ && - X_{i}N_e(\alpha_{i \rightarrow i-1}(\tilde{\varepsilon}) + C_{i \rightarrow i+1}(\tilde{\varepsilon})) \label{ionbaleqn}
\end{eqnarray}
where $X_i$ is the total population in the $i^{th}$ ionisation stage, $N_e$ the electron number density, $\tilde\varepsilon$ a variable that describes the shape of the electron distribution (in the case of this work $\tilde\varepsilon$ is the electron temperature as all results presented are for thermal systems with a Maxwellian distribution), $C$ is a bulk collisional ionisation rate coefficient, $\beta$ a bulk 3-body recombination rate coefficient, and $\alpha$ a bulk recombination rate coefficient (which includes radiative and dielectronic recombination).  Because all results presented in this work neglect the effect of a radiation field, photoionization and stimulated recombination have been omitted from equation (\ref{ionbaleqn}).  While GSM is capable of calculating these bulk rates by integrating the raw level-specific cross sections and summing over all intermediate pathways, in general that approach would require a data set that is significantly larger than desired.  As such these rates, when available, come from either tabulated or fitted data in the literature (such as can be found in \citet{Arnaud-Rothenflug:1985,Arnaud-Raymond:1992} or \citet{Mazzotta-etal.1998}).  All of the calculations presented in this study are steady-state calculations and therefore the left-hand side of equation (\ref{ionbaleqn}) can be set to zero.  We note that GSM is also explicitly designed to handle transient sources.

Once the total ionisation stage populations have been calculated, they are used to calculate the individual level populations, and subsequently the spectra, by solving the following set of coupled equations:
\begin{eqnarray}
\frac{dN_{l,j}}{dt} & = &N_e\left(X_{l-1}\alpha^{\mathrm{eff}}_{l-1,1 \rightarrow l,j}(\tilde{\varepsilon}) -N_{l,j}\sum_{i}\alpha^{\mathrm{eff}}_{j,l \rightarrow i,l+1}(\tilde{\varepsilon})\right.\nonumber \\
&+& X_{l+1}C^{\mathrm{eff}}_{l+1,1 \rightarrow l,j}(\tilde{\varepsilon})-N_{l,j}\sum_{i}C^{\mathrm{eff}}_{l,j \rightarrow l-1,i}(\tilde{\varepsilon}) \nonumber \\ 
&+&\left.\sum_{i,i\ne j}(N_{l,i}q^{\mathrm{eff}}_{i \rightarrow j}(\tilde{\varepsilon})-N_{l,j}q^{\mathrm{eff}}_{j \rightarrow i}(\tilde{\varepsilon}))\right)\nonumber \\ 
&+&N_e^2\left(X_{l-1}\beta^{\mathrm{eff}}_{l-1,1 \rightarrow l,j}(\tilde{\varepsilon})-N_{l,j}\sum_{i}\beta^{\mathrm{eff}}_{j,l \rightarrow i,l+1}(\tilde{\varepsilon})\right) \nonumber \\
 &+&\sum_{i,i>j}N_{l,i}A^{\mathrm{eff}}_{i \rightarrow j} -N_{l,j}\sum_{i,i<j}A^{\mathrm{eff}}_{j \rightarrow i} \nonumber \\
&-&N_{l,j}\sum_{i} R^{\mathrm{AI-eff}}_{l,j \rightarrow i,l-1}
 \label{excitedformula}
\end{eqnarray}
where the variables are defined more or less as before, with $N_{l,j}$ being the population in the $j^{th}$ state of the $l^{th}$ ionisation stage, $q^{\mathrm{eff}}$ an electron-impact (de-)excitation effective rate coefficient, $R^{\mathrm{AI-eff}}$ an autoionization rate coefficient, and $A^{\mathrm{eff}}$ an effective radiative decay rate.  Just as in equation (\ref{ionbaleqn}), equation (\ref{excitedformula}) omits terms that involve the radiation field.  It also omits excitation by proton and $\alpha$ particle impact, as they are neglected in this study for the sake of simplicity in order to better illustrate the effect of level-specific recombination rates on the $G$ ratio.  Once again all calculations consider steady-state conditions, so the left-hand side of equation (\ref{excitedformula}) can be set to zero.

As mentioned earlier, GSM does not treat every spectroscopic level the same.  Instead, it divides them into two categories: explicit (which are included as levels in equation (\ref{excitedformula}), and coupled via all processes considered), and statistical (the effects of which are included when calculating the rate coefficients that go into equation (\ref{excitedformula})).  The inclusion of statistical states in the rate coefficients is accomplished via the collisionless transition matrix (CTM).  A given element of the CTM can be thought of as the probability that once a statistical state gets populated, neglecting collisions, the atom or ion will end up in a given explicit state without passing through any other explicit states.  If $Q$ is the set of explicit states, the CTM can be defined using the following recursive expression:
\begin{eqnarray}
\mathbf{T}_{i\rightarrow j}&=&\sum\limits_{k\not\in Q \atop (E_i>E_k>E_j)}\frac{\Gamma_{i\rightarrow k}}{\sum\limits_lA_{i\rightarrow l}+\sum\limits_m R^\mathrm{AI}_{i \rightarrow m}}\mathbf{T}_{k\rightarrow j}\nonumber \\
&+&\frac{\Gamma_{i\rightarrow j}}{\sum\limits_lA_{i\rightarrow l}+\sum\limits_m R^\mathrm{AI}_{i\rightarrow m}}
\end{eqnarray}
where $\Gamma_{i\rightarrow k}$ is the appropriate type of spontaneous rate to connect $i$ and $k$, via either radiative decay or autoionization.  It should be noted that in order for this quantity to be meaningful, the state $i$ must not be in the set $Q$.

After calculating the CTM, each effective rate coefficient can be constructed from the direct rate coefficient connecting the two explicit states and sums over all the indirect paths.  For example effective collisional ionisation rate coefficients are calculated as
\begin{eqnarray} 
C^{\mathrm{eff}}_{i\rightarrow j}(\varepsilon) &=& C^{\mathrm{direct}}_{i\rightarrow j}(\varepsilon) + \sum_{k\atop (E_k>E_j)}C^{\mathrm{direct}}_{i\rightarrow k}(\varepsilon)\mathbf{T}_{k \rightarrow j} \nonumber \\ &+&\sum_{m \atop (E_m>0)}q^{\mathrm{direct}}_{i \rightarrow m}(\varepsilon)\mathbf{T}_{m \rightarrow j} \label{cieq} \;,
\end{eqnarray}
recombination (RR+DR) rate coefficients as
\begin{eqnarray}
\alpha^{\mathrm{eff}}_{j \rightarrow i}(\varepsilon) &=& \alpha^\mathrm{RR}_{j \rightarrow i}(\varepsilon) + D^\mathrm{DC}_{j \rightarrow i}(\varepsilon) + \sum_{m\atop(E_m>E_i)}\alpha^\mathrm{RR}_{j \rightarrow m}(\varepsilon)\mathbf{T}_{m \rightarrow i} \nonumber \\ &+& \sum_{m \atop (E_m>E_j,E_m>E_i,E_m>0)}D^\mathrm{DC}_{j \rightarrow m}(\varepsilon)\mathbf{T}_{m \rightarrow i} \label{rceq}\;,
\end{eqnarray}
where $D^\mathrm{DC}$ is a dielectronic capture rate, and electron-impact excitation and de-excitation rate coefficients are calculated as
\begin{eqnarray}
q^{\mathrm{eff}}_{j \rightarrow k}(\varepsilon) & = & q^{\mathrm{direct}}_{j \rightarrow k}(\varepsilon) + \sum_{l \atop (E_l>E_j,E_l>E_k)}q^{\mathrm{direct}}_{j \rightarrow l}(\varepsilon)\mathbf{T}_{l \rightarrow k} \nonumber \\
&+&  \sum_{i \atop (E_i>E_j,E_i>E_k,E_i>0)}D^\mathrm{DC}_{j \rightarrow i}(\varepsilon)\mathbf{T}_{i \rightarrow k}\label{eieeq} \;.
\end{eqnarray}
It should be noted that in equations (\ref{cieq})--(\ref{eieeq}) the effect of resonances are included as indirect pathways via terms involving dielectronic capture ($D^\mathrm{DC}$).  When R-Matrix data are used, these terms are omitted to avoid double counting the effect of resonances.

Once equation (\ref{excitedformula}) has been solved and the level populations obtained, synthetic spectra are calculated.  The first step is to calculate the intensity (in number of photons cm$^{-3}$s$^{-1}$) of all lines in the desired energy range as
\begin{equation}
I(l,j\rightarrow l,k)=N_{l,j}A_{j \rightarrow k}\;.
\end{equation}
Each line is then given a line shape corresponding to a thermal Doppler-broadened Gaussian profile.  The total spectra, $S$, for a given photon energy, $h\nu$, can be expressed as
\begin{equation}
S(h\nu)=\sum_s I_sh\nu\frac{c\sqrt{m_i}}{2\pi kT}e^{\frac{m_ic^2(h\nu-\Delta E_{ij})^2}{2\Delta E_{ij}^2kT_i}}\;,
\end{equation}
where $s$ ranges over the set of all included transitions in the desired energy range, and the ion temperature, $T_i$, is taken to be equal to the electron temperature.

\section{Computations}
The baseline model considered in the present work uses the ionisation balance data of \citet{Mazzotta-etal.1998} and detailed, level-specific data calculated with the Los Alamos suite of atomic physics codes \citep[see for instance][]{Abdallah-etal.1994,Abdallah-etal.2001}.  The first step in such a calculation is to use the CATS code to calculate the bound and autoionizing wave functions, energies, and dipole allowed radiative decay rates for all fine-structure levels arising from the configurations $\mathrm{nl}$, $\mathrm{1snl}$, $\mathrm{2lnl'}$, $\mathrm{1s^2nl}$, $\mathrm{1s2lnl'}$, and $\mathrm{1s3lnl'}$ with $\mathrm{n}\le 10$ and $\mathrm{l}\le g$ which span the H-like, He-like, and Li-like ionisation stages.  Only fine-structure levels arising from the $\mathrm{1s}$, $\mathrm{1s^2}$, $\mathrm{1s2l}$, $\mathrm{1s^22l}$, and $\mathrm{1s2lnl'}$ configurations with $\mathrm{n}\le 10$ and $\mathrm{l}\le g$ were treated explicitly.  All other levels were treated statistically.  Convergence with respect to $\mathrm{n}$ was tested with a larger model that included up to $\mathrm{n}\le 20$.  The GIPPER code is then used to calculate autoionization and photoionization cross sections in the distorted-wave approximation, as well as collisional ionisation cross sections.  As the ions under consideration are highly charged, a scaled hydrogenic approximation was used for collisional ionisation cross sections as it agrees well with distorted-wave results for systems such as these and is much more computationally efficient.  The ACE code was used to calculate distorted-wave cross sections for all electron-impact excitation transitions out of the lowest seven levels of the helium-like ionisation stage, as well as the $\mathrm{1s^22l}$ complex of the Li-like ionisation stage (which is used to obtain effective collisional ionisation rate coefficients into the He-like stage as well as a method for producing the K$\alpha$ satellite lines).  Lastly, the non-dipole $A$ values that give rise to the `x' and `z' lines as well as a two-photon decay rate from $\mathrm{1s2\;^1S_0 \rightarrow 1s^2\;^1S_0}$ were obtained from \citet{Mewe-Schrijver:1978stat}. All rates were calculated from these cross sections inside of GSM.  We refer to this baseline model as DW.  

The second model, which is designated RM, begins with the first model and replaces the radiative and dielectronic recombination data from the H-like ionisation stage into the He-like ionisation stage with total and level-specific rate coefficients for recombination into all fine-structure levels SLJ with $\mathrm{n}\le10$  calculated from Breit-Pauli R-Matrix photoionization cross sections \citep{Nahar-Pradhan:1994-urr}.  While there are unified recombination data available in the literature for Ni \citep{Nahar:2005} and Fe \citep{Nahar-etal:2001-FeXXIV-FeXXVrc}, we do not consider a Ca, RM model as there are no data yet available for this species.  In addition, the energy levels and radiative decay rates inside the He-like ion are replaced by R-Matrix data (the radiative decay rates are taken from Nahar \& Eissner (in preparation) for Ni and \citet{Nahar-Pradhan:1999} for Fe).  Lastly, the unified recombination rates through a given satellite line of \citet{Nahar-sat-rc} are used to calculate the dielectronic recombination contribution to the intensity of all the KLL satellite lines in this model.

The third model, referred to as M85, also begins with the baseline model and replaces the recombination rates from the H-like ionisation stage into the He-like excited states with rates calculated using the expressions of \citet{Mewe-etal.1985}.  These expressions are fits to previously published data.  In the case of Ni, the fits are to the rates of \citet{Mewe-Gronenschild:1981}, which are based on the work of \citet{Burgess-Seaton-1960} concerning the recombination of neutral He.  These fits do not include the contribution of dielectronic recombination, so in the Ni M85 model dielectronic recombination is included using the data from the baseline DW model.  \citet{Mewe-etal.1985} fit the data of \citet{Bely-Dubau-etal.1982a} for Fe and \citet{Bely-Dubau-etal.1982b} for Ca.  Both of these works are based on the hydrogenic recombination work of \cite{Burgess:1958}.  It should be noted that \citet{Bely-Dubau-etal.1982a} and \citet{Bely-Dubau-etal.1982b} only present rates up to an electron temperature of $10^8$~K.  As such, using the fits of \citet{Mewe-etal.1985} to these data beyond this temperature may exceed the range of validity intended by \citet{Bely-Dubau-etal.1982a,Bely-Dubau-etal.1982b}.  Furthermore these rates do include the effect of dielectronic recombination, so this contribution is not recalculated and included a second time by GSM.

The fourth model, MS78, is much like the third, but instead calculates radiative recombination rates using the expressions of \citet{Mewe-Schrijver:1978stat}, which are also based on the work of \citet{Burgess-Seaton-1960}.  As these rates do not include dielectronic recombination, this model includes DR rates obtained from the data contained in the baseline, DW model.

\section{Results}

\begin{figure*}
\resizebox{!}{0.695\textheight}{\includegraphics{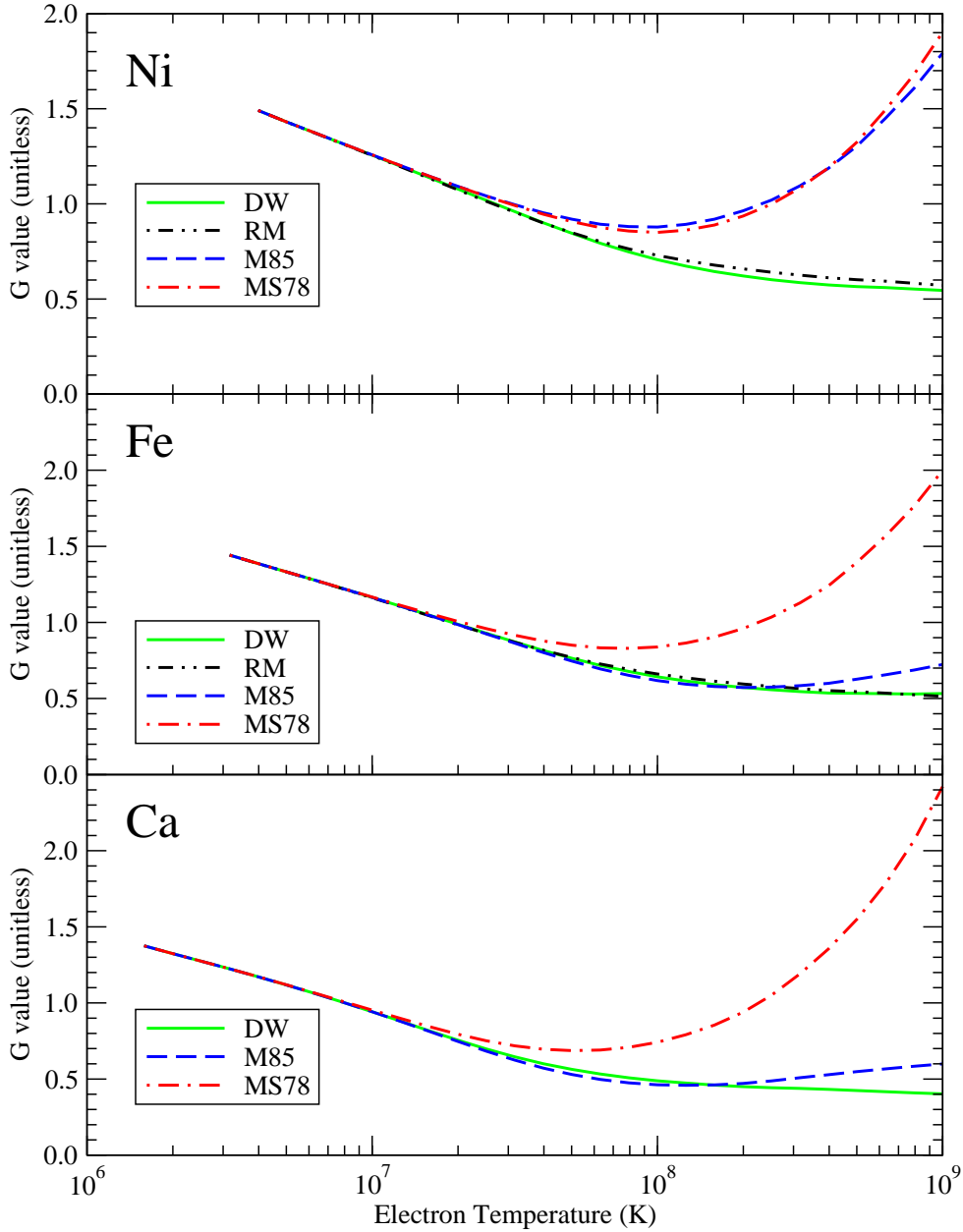}}
\caption{$G$ ratios for Ni (top), Fe (middle) and Ca (bottom) for each of the models considered.\label{G-fig}}
\end{figure*}

The results for the $G$ ratio for all models under consideration are presented in figure \ref{G-fig}.  The differences are striking.  As temperature increases and recombination becomes more important to spectral formation, the more approximate rates of \citet{Mewe-Schrijver:1978stat} produce $G$ ratios that diverge significantly from the results obtained with the newer DW and RM models for all three elements.  The M85 result for Ni displays similar discrepancies while the M85 results for Fe and Ca (which are fits to \citet{Bely-Dubau-etal.1982a,Bely-Dubau-etal.1982b}) agree better with the RM and DW models.  Not surprisingly (given the highly charged nature of these systems) the RM and DW models are in excellent agreement at all temperatures.

\begin{figure*}
\resizebox{!}{0.695\textheight}{\includegraphics{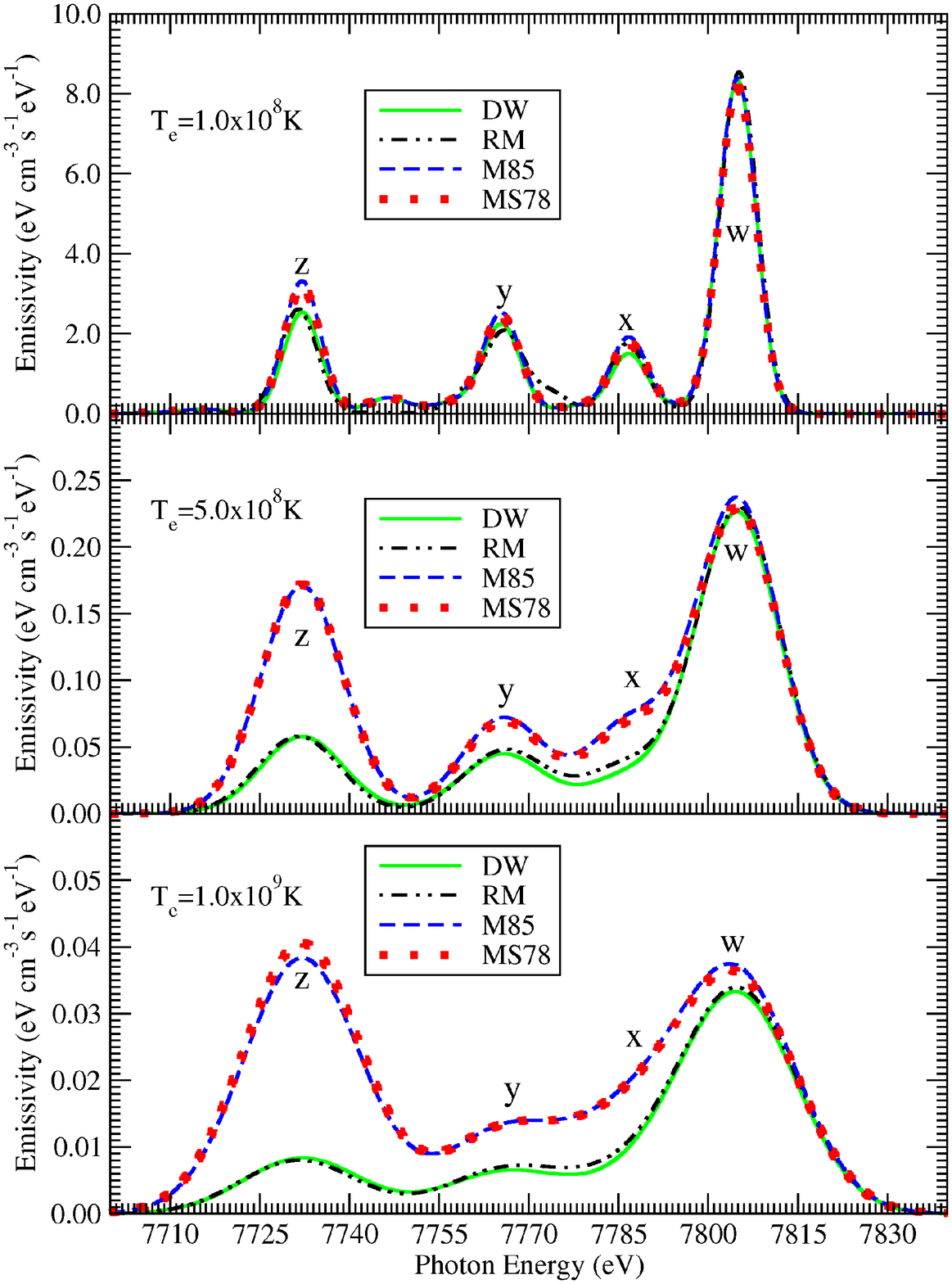}}
\caption{Emission spectra of a low density ($N_e=10^{10}$cm$^{-3}$) Ni plasma at three different temperatures in the high-temperature range where the $G$ ratio from the models is in poor agreement.  Spectra from the DW, M85, and MS78 models have all been red shifted by 9 eV to facilitate an easier comparison.\label{Ni-spec}}
\end{figure*}

\begin{figure*}
\resizebox{!}{0.695\textheight}{\includegraphics{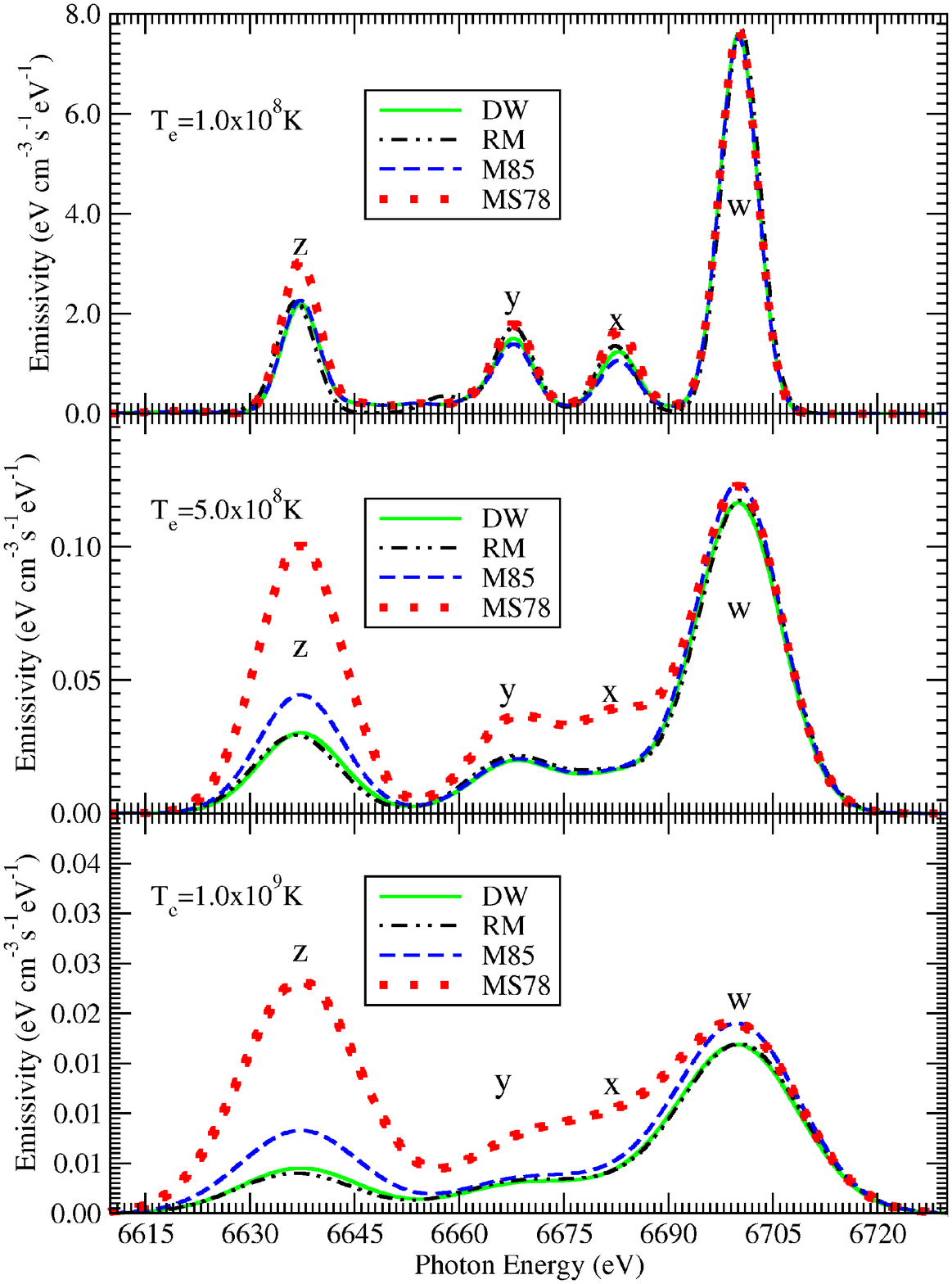}}
\caption{Emission spectra of a low density ($N_e=10^{10}$cm$^{-3}$) Fe plasma at three different temperatures in the high-temperature range where the $G$ ratio from the models is in poor agreement.  Spectra from the DW, M85, and MS78 model have all been red shifted by 7 eV to facilitate an easier comparison.\label{Fe-spec}}
\end{figure*}

\begin{figure*}
\resizebox{!}{0.695\textheight}{\includegraphics{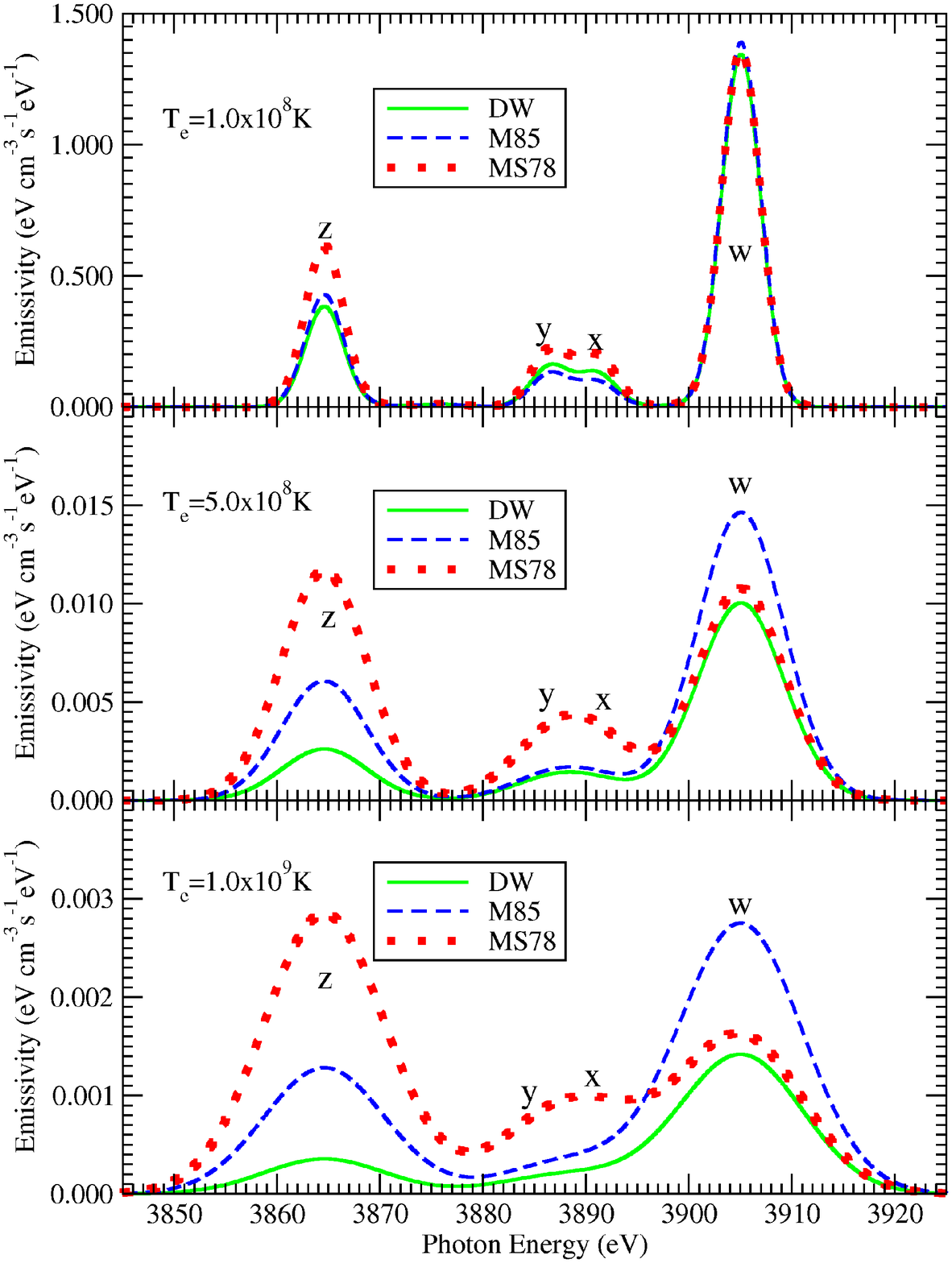}}
\caption{Emission spectra of a low density ($N_e=10^{10}$cm$^{-3}$) Ca plasma at three different temperatures in the high-temperature range where the $G$ ratio from the models is in poor agreement.\label{Ca-spec}}
\end{figure*}

Upon investigation of the underlying spectra (figures \ref{Ni-spec}--\ref{Ca-spec}) to find the source of the divergence in predicted $G$ values at high temperature, one continues to see excellent agreement at all temperatures between the RM and DW models; but the analysis is not as straightforward for the more approximate methods.  In the case of Ni (figure \ref{Ni-spec}), both the M85 and MS78 model agree well with each other, predicting significantly more flux in the `x', `y', and `z' lines, and slightly more flux in the `w' line, at high temperatures than either of the more accurate models.  This behaviour is consistent with the fact that both models are based on the same underlying work by \cite{Burgess-Seaton-1960}.  The spectrum for Fe (figure \ref{Fe-spec}) is more interesting because, while the MS78 model continues to predict significantly stronger `x', `y', and `z' lines, and a slightly stronger `w' line, compared to the DW and RM models, the M85 model predicts a rather different result. Like the MS78 model, the M85 model predicts a slightly stronger `w' line, but the `z' line is in much better agreement with the DW and RM models than the MS78 model.  Furthermore the M85 model does a good job of predicting the `x' and `y' line intensities. 

The detailed results for Ca (figure \ref{Ca-spec}) are somewhat similar to those presented for Fe.  The MS78 model continues to predict much stronger `x', `y', and `z' lines when compared to the DW data set and only a slightly stronger `w' line.  The M85 model is in good agreement for the `x' and `y' lines and much closer for the `z' line, but in the case of Ca it predicts a much stronger `w' line.  This over prediction tends to mitigate the effect of the enhanced `z' line, producing a fortuitously lower value for the $G$ ratio.  This type of cancellation underscores the importance of examining the actual spectra to gain a better understanding of the quality of the underlying atomic data.

\begin{figure*}
\resizebox{!}{0.695\textheight}{\includegraphics{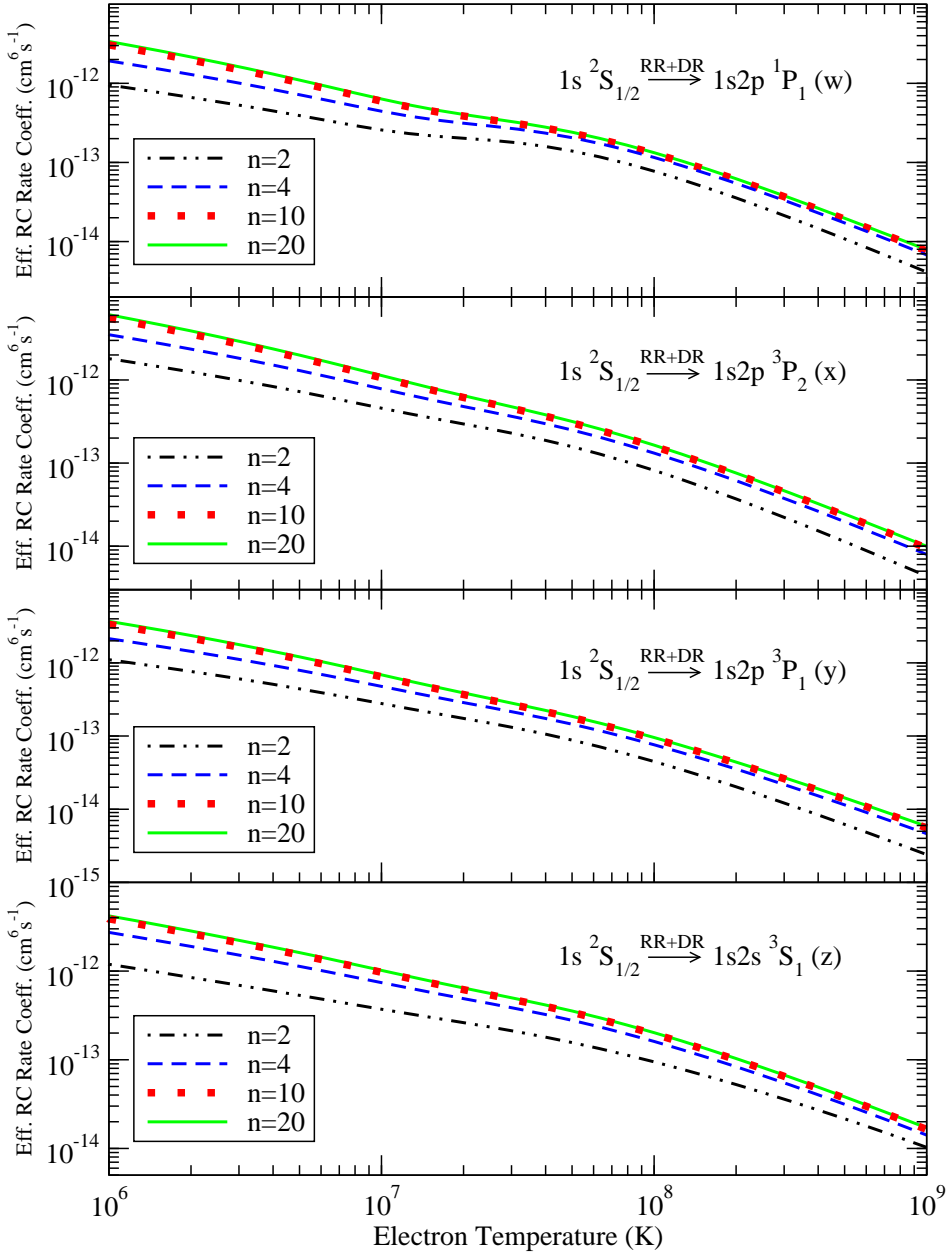}}
\caption{Four effective recombination rate coefficients for Fe as a function of the highest principal quantum number included in the DW model.  Of particular interest is how quickly the rate coefficient converges in the high-temperature region due to cascade terms becoming less important.\label{Fe-rc-conv}}
\end{figure*}

In order to determine the fundamental reasons behind the above discrepancies in the spectra and $G$ ratios, the recombination rates must be examined in detail, but first it is useful to ensure that the models are large enough (as a function of the principal quantum number $\mathrm{n}$) to have converged upon the correct answer.  As the various sources cited in this work all chose different limits for the maximum $\mathrm{n}$ value used in their calculations, lack of convergence could be a reason for the differences.  Additionally, such a study can be used to investigate the relative importance of cascade corrections, which \citet{Pradhan-Shull:1981-helines} suggested to be the cause of the upturn in the $G$ ratio displayed in that work.  As the results for each element are very similar, only recombination rate coefficients for Fe are presented (figure \ref{Fe-rc-conv}) and discussed for only the DW model.
 
\begin{figure*}
\resizebox{!}{0.695\textheight}{\includegraphics{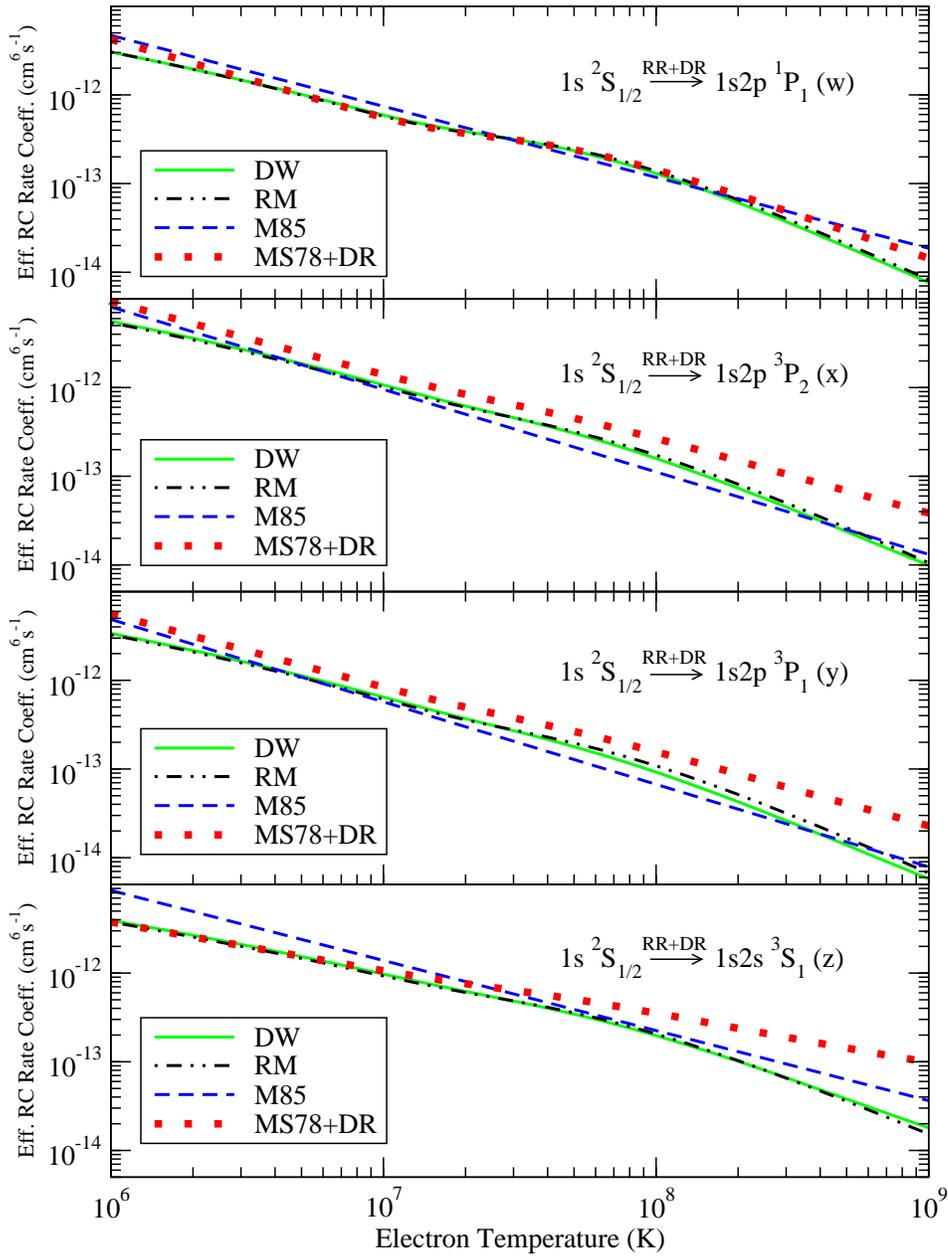}}
\caption{Comparison of the effective recombination rate coefficients for Fe for the various models.  Of particular note are the large discrepancies at electron temperatures above $10^8$~K.\label{Fe-rc}}
\end{figure*}

In figure \ref{Fe-rc-conv} it can be seen that by $\mathrm{n}=10$ the effective rate coefficients into the relevant four spectroscopic levels have converged to within $\sim$5\ per cent when the temperature is above $10^8$~K.  All models considered in this work extend at least this far in principal quantum number; thus, the extent of the model (in $\mathrm{n}$) is not the cause of the discrepancies.  Of additional note is the fact that figure~\ref{Fe-rc-conv} indicates that recombination cascades become less important (as a method of populating the He-like excited states) at higher temperatures, even as recombination is becoming a more important process for the population of He-like excited states.  According to the analysis of the DW model, it is the direct recombination rate that is the most important process at higher temperatures while cascades play a progressively smaller role.

Moving on to a comparison of the effective recombination rate coefficients in the four models (again presented only for Fe since the Ca and Ni results show similar trends) shown in figure \ref{Fe-rc}, there are obvious discrepancies at high temperatures and the cause of the differences in simulated spectra becomes evident.  The two, more approximate models (MS78 and M85) are in reasonable agreement with the more accurate DW and RM models in the middle-temperature region.  Outside the range, there is a marked difference (as much as a factor of five in the high-temperature region).  This difference in the recombination rate coefficients in the high-temperature regime is responsible for the corresponding difference in the $G$ ratio among the various calculations.  The values of the new DW and RM recombination data are significantly smaller than the MS78 data, which result in decreased recombination rates for populating the `x', `y', and `z' lines, while the impact-excitation rate for feeding the `w' line remains essentially fixed for each type of calculation. Thus, the reduced DW and RM recombination data cause a delay in the onset of any increase in the $G$ ratio to a higher temperature, resulting in the large differences displayed in figure \ref{G-fig}.

It should be noted that \citet{Nahar-etal:2001-FeXXIV-FeXXVrc} saw a similar difference in the recombination rates in the high-temperature region when they compared their level-specific unified recombination rates with the direct component of the recombination rate expressions of \citet{Mewe-Schrijver:1978stat}.  Again there is excellent agreement between the DW and RM models over the entire temperature range, bolstering confidence in their validity.

\section{Conclusion}
While the results of calculations using recombination rates derived from more accurate and extensive distorted-wave and R-Matrix calculations agree quite well, both disagree significantly with the previous, more approximate calculations of \citet{Mewe-Schrijver:1978stat,Mewe-Gronenschild:1981,Bely-Dubau-etal.1982a,Bely-Dubau-etal.1982b} and \citet{Mewe-etal.1985}.  The results obtained from models based on the more accurate distorted-wave and R-Matrix data show a continually decreasing $G$ ratio over the investigated temperature range.  This behaviour disagrees strongly with the sharp upturn observed at high temperatures in the $G$ ratios calculated from the more approximate models.  The discrepancy is up to a factor of six in the $G$ ratio (see figure \ref{G-fig}), and was determined to be caused by large differences in the recombination contributions.  The more approximate rates produce results that agree reasonably well with the newer DW and RM data up to approximately $10^8$~K, but can lead to unphysical values when used outside of this range.

Further work is required to refine and verify calculations of the $G$ ratio and the underlying data from which it is determined.  Studies similar to the present one are needed, but including electron-impact excitation rates as well as collisional ionisation rates from the ground state of the Li-like species into the excited states of the He-like species.  Investigation of the effect of the ionisation balance may also be required.  These studies are underway.

\section{acknowledgments}
This work was partially conducted under the auspices of the United States Department of Energy at Los Alamos National Laboratory.  Much of the development of GSM was also done at the Ohio Supercomputer Center in Columbus, Ohio (USA).  The work by the OSU group (MM, SNN, AKP) was partially supported by a grant from the NASA Astrophysical Theory Program. 

%\bibliography{references}

\label{lastpage}
\end{document}